\documentstyle[prl,aps]{revtex}

\def\textbf#1{{\bf #1}}
\def\be{\begin{equation}}
\def\ee{\end{equation}}
\def\ben{\begin{eqnarray}}
\def\een{\end{eqnarray}}
\def\eea{\end{array}}
\def\bea{\begin{array}}
\newcommand{\bei}{\begin{itemize}}
\newcommand{\eei}{\end{itemize}}

\begin{document}
\draft
\twocolumn

\title{
Mean of continuous variables observable {\it via}
measurement on single qubit
}

\author{Pawe\l{} Horodecki}

\address{Faculty of Applied Physics and Mathematics,
Technical University of Gda\'nsk, 80--952 Gda\'nsk,
Poland}
\maketitle

\begin{abstract}
It is shown that mean value of any observable with bounded 
spectrum can be uniquely determined from binary statistics of 
the measurement performed on {\it single} qubit ancilla
coupled to a given system. The observable structure is fully encoded 
in the corresponding POVM. The method is generalised to the case of
distant labs paradigm and discussed in the context  of entanglement 
detection with few local measurements. The results  are also discussed in 
the context of quantum programming.
\end{abstract}

One of the serious problems of quantum information
theory \cite{Ek,Nielsen,Springer,Gruska} is  
fragility of quantum entanglement and other fully quantum 
resources (like quantum gates) that are basic 
ingredient form most of quantum information phenomena \cite{QIT}.

However before using entanglement or other resource, one has to be sure that
it really is present in the system (see \cite{Jaynes} for some paradoxes).
This domain involves seriously quantum measurement theory.

In particular it is important to be able to detect
entanglement in distant labs scenario, where two observers
are far apart and have restricted access to the
composed system they share (see for instance
\cite{QOV}). There are many methods checking whether there
is entanglement in the system (see \cite{Springer,JModOpt}).
However they require prior state reconstruction ie.
full knowledge about density matrix of the system.
Recently new paradigm was introduced \cite{PHAE}
in which one asks about entanglement of unknown 
state produced by stationary source.
The answers has been given in series of papers 
\cite{PHAE,PH,Guhne}. In particular for
two qubits entanglement can be detected
both qualitatively \cite{PHAE} and quantitatively \cite{PH}
without state reconstruction.
If partial information about the state is provided, then entanglement
can be detected \cite{Guhne} in distant labs paradigm
with minimal number of product observables locally estimated.

The problem is that, according to quantum measurement theory,
even determining of mean value of {\it single} (sic) observable 
$A$ requires typically estimation of
{\it many} parameters namely the probabilities of outcomes
of von Neumann measurement (see {\it Problem} below).
The problem is especially striking in continuous variables case
\cite{CV} where any von Neumann measurement can be only approximate
due to finite number of outcomes of {\it any} real experiment.
In this context we address quite general:

{\it Question. - Is there any way to determine mean value of given
quantum observable $A$ from experimental estimation of single parameter ?}

Here by estimation of single parameter we mean estimation
of probability of some single outcome ie. 
the number of clicks of single detector divided by the 
number of all runs of experiment.  The simplest example is spin
polarisation measurement along given axis:
to get probability of being ``up'' we count ``up''
events and divide them by the number of all
(``up'' and ``down'') results.

Surprisingly, the answer for the Question
above is positive for {\it any bounded observable,
no matter whether it involves continuous variables or not}.
The nature of the associated effect seems to be quite fundamental,
and it not been known in quantum measurement theory so far.

It can be explained as follows: if apart form our system
we have {\it single qubit} and can control the
system-qubit interaction then there exists general
quantum measurement (POVM) with two outcomes such that
mean value of $A$ can be immediately reconstructed from
the POVM statistics. Because binary POVM corresponds
to estimation of single parameter (see previous discussion)
it happens that in the presented scheme estimation of
mean value of single observable does correspond
to single parameter.
The mechanism of this effect can be roughly summarised
with the statement that the observable has been {\it encoded} into
the interaction (represented by POVM) between the system and
the qubit ancilla.

In context of results of Ref. \cite{Guhne} we also
pose similar issue in case of product observables
measured by distant observers.
It happens again that two binary POVM-s are sufficient
but with data analysis refined to get apart form
marginal statistics also one correlation probability
(like in Bell-type experiments).

Let us note that as a byproduct of other
investigations we have provided partial
positive answer to the above Question \cite{Estimator}.
The idea was to encode any spin-like observable $A$ into some state
$\varrho_{A}=\alpha I + \beta A$ (that
can be viewed as a kind of program) of auxiliary system.
Then the mean value of $A$ in given $\varrho$ was determined  
from the value $Tr(\varrho_{A}\varrho)$
that was found to be easily measurable as a single parameter
in  some interferometric scheme \cite{Estimator}.
However that scheme required complicated resources:
any d-level system needs $(d+1)$-level ancilla.
Moreover, as discussed subsequently, it can not be
applied for continuous variables case \cite{CV}.

Here we provide novel unified approach that has quite surprising 
advantages:

(i) requires minimal ancilla - just single quantum bit

(ii) is applicable for continuous variables case
under the only assumption of boundness of the observable.

Furthermore the approach allows for very easy extension to LOCC schemes.
The present analysis has also new general motivation: need of 
many parameters estimation  in standard von Neumann measurement.
As we already mentioned the LOCC scheme we provide here
is especially important for local detection of
unknown or partially unknown entanglement. In particular
it provides additional justification to approaches
from  Ref. \cite{Guhne}.

It is worth to mention that recently one developed
ideas of quantum computing with quantum 
data structure \cite{Estimator,PHAE},
quantum programmable interferometric networks
\cite{Estimator} or programmable quantum gates \cite{gates}.
In this context we address natural (open) question:
what observable can be implemented as a kind of quantum
program, and, if so, how to do it in optimal way and how to
quantify this process.

The paper is organised as follows. First
we pose the problem with standard von Neumann
measurement and we show how solve it
encoding given observable into binary POVM.
Then we provide similar
result for product observable in LOCC paradigm.
Finally we briefly discuss the result especially
in context of recently considered computing with
quantum data structure and related issues.

{\it The Problem .-}
Consider arbitrary quantum observable $A=\sum_{i}\lambda_{i}
| \psi_{i}\rangle \langle \psi_{i}|$. If it has more than two different
eigenvalues $\lambda_{1}$, $\lambda_{2}$, ..., $\lambda_{n}$, $n>2$
then usual procedure to get mean value of $A$ in given state
$\varrho$
\begin{equation}
\langle A \rangle_{\varrho}=Tr(A \varrho)
\end{equation}
requires von Neumann measurement with $n$ outcomes
corresponding to $n$ eigenvectors of the state $\varrho$:
$\psi_{1}$, $\psi_{2}$, ..., $\psi_{n}$.
The measurement rely on estimation of $n-1$ parameters
that are probabilities of outcomes
$p_{1}=\langle \psi_{1}| \varrho | \psi_{1} \rangle$,
$p_{2}=\langle \psi_{2}| \varrho | \psi_{2} \rangle$,...,
$p_{n-1}=\langle \psi_{n-1}| \varrho | \psi_{n-1} \rangle$
(the last parameter $p_{n}$ can be inferred from normalisation condition).
Finally we multiply probabilities by eigenvalues
and calculate the sum $\sum_{i=1}^{n} p_{i} \lambda_{i}$
which is equivalent just to the mean value
$ \langle A \rangle_{\varrho}$ we were looking for.
Clearly if $n$ is greater than two we
need estimation of more than one parameter in the sense
that (apart from counting runs of our experiment)
we have to count clicks corresponding
to  {\it more than one} outcomes.
Moreover if the observable correspond to continuous
variables case ($n=\infty$ above)
than there is no way to measure it directly and any
indirect measurement must be approximate.

We shall see however that one can overcome those
disadvantages under two assumptions:
(i) boundness of the observable spectrum
(ii) additional resource: well controlled interaction
with single quantum bit.

{\it Solution for single observable .-}
Let us first assume now that the observable
has the spectrum bounded and its lower and upper
bound correspond to $a_{min}$ and $a_{max}$ respectively.
Let us define the non negative number $a_{-}\equiv min[0,-a_{min}]$.
Then the following operator $D=a_{-}I+A$ is positive
($D \geq 0$) ie. has no negative eigenvalue.
Now we define the new operator hermitian operator
$D'=D/(a_{+})$ ($a_{+}\equiv max[a_{-},a_{-}+a_{max}]$)
satisfies the property $0\leq D' \leq 1$
\footnote{We say that $A \geq B$ if for all $\Psi$
one has $\langle \Psi|A-B|\Psi\rangle \geq 0$.}
which means that it all eigenvalues belong to the interval
$[0,1]$. Consider now the following operators
\begin{eqnarray}
&&V_{0}=\sqrt{D'}=\sqrt{(a_{-}I+A)/a_{+}}, \nonumber \\
&& V_{1}=\sqrt{I - V_{0}^\dagger V_{0}}.
\end{eqnarray}
They satisfy the following condition $\sum_{i=0}^{1}V_{i}^{\dagger}V_{i}=I$
so they represent so called generalised quantum measurement and can be
easily implemented on our system. It has two outcomes $i=0,1$
with probabilities $p_{0}=Tr(V_{0}^\dagger V_{0}\varrho)$,
$p_{1}=Tr(V_{1}^\dagger V_{1}\varrho)=1-p_{0}$. Note that
only {\it single} parameter $p_{0}$ describes this binary statistics.
Now it is elementary to see that because of hermicity of $V_{0}$
(which means that $V_{0}=V_{0}^{\dagger}$) one has
$p_{0}=Tr((a_{-}I+A)\varrho)/a_+$ and finally, because
of $Tr(A\varrho)=\langle A\rangle_{\varrho}$ this leads to
the main conclusion
\begin{equation}
\langle A \rangle_{\varrho}=a_{+} p_{0} - a_{-}.
\end{equation}
Thus we have reproduced mean value of arbitrary observable
$A$ with bounded spectrum with help of
single parameter $p_{0}$ coming from binary generalised
quantum measurement (POVM).

It is remarkable that the above POVM can be performed
on the system if only we have
one qubit ancilla (additional physical system)
and can control interaction between our system and the
ancilla. Indeed this is all what binary POVM requires
(\cite{Kraus}, see Appendix
of \cite{QOV} for tutorial review). We prepare our
ancilla qubit in the pure state $|0\rangle$. Then
we perform unitary evolution $U$ of our global
system: ``system in state $\varrho$ + ancilla''.
Finally we measure observable $\sigma_z$ on our ancilla.
If we get result ``up'' (ancilla state unchanged ie.
remains in initial $|0\rangle$)
this corresponds to result $i=0$,
if we get result ``down'' (ancilla state changed
into $|1\rangle$) this corresponds to the
result $i=1$ both occurring with probabilities
$p_{0}$, $p_{1}$ defined above.

At a first glance the present scheme has close similarity to that of 
universal quantum estimator detecting nonlinear
state functions \cite{Estimator} as in the latter
only  single control qubit is finally measured.
There is a significant difference however, 
that makes the present scheme successful for CV case.
We shall discuss that issue in the conclusion part 
of the paper.

Note that the above scheme allows to detect ``mean value'' of
nonhermitian operator $X$ defined by $\langle X \rangle_{\varrho}$
with help of decomposing $X$ into hermitian and antihermitian part
( cf. \cite{PHSPA}) and detecting the corresponding observables with
help of two binary POVM-s.


{\it Product observables and distant labs paradigm .-}

Suppose now that Alice and Bob are in distant labs
paradigm ie.  they are far apart ad they share
some bipartite quantum state $\varrho_{AB}$.
This is like in quantum teleportation
process where they shared single state
(here we allow $\varrho$ to be mixed).
In such case Alice and Bob are allowed to perform
local operations (LO) and communicate classically (CC).
Suppose now that they want to detect
mean value of some entanglement witness
$W=\sum_{k=1}^{m} A_{k} \otimes B_{k}$ with its structure
and number chosen properly (see \cite{Guhne}).
Because of LOCC restrictions this can be achieved
only by measurement of local measurements
and exchange of information. Usually it is done like
in standard Bell inequalities
(for similarity of entanglement witnesses formalism
to Bell inequalities theory see \cite{TerhalBell}).
Namely this corresponds to local measurements of
observables $A_{k}$, $B_{k}$ (for each fixed $k$)
but with keeping the record of results and finally
establishing the mean value form joint statistics.
However there is more outcomes in general so again there is
a question  whether we can reduce the above
scheme to binary experiments.
The answer is ``Yes'', though the solution
is not so simple as it was before.
Suppose that Alice and Bob want to measure mean value
\begin{equation}
\langle A \otimes B\rangle_{\varrho_{AB}}=
Tr(A \otimes B \varrho_{AB})
\end{equation}
of product observable $A \otimes B$
on shared state $\varrho_{AB}$.
Then they should perform local POVM-s corresponding
to local observables as defined in previous section but
they should use the data in more sophisticated way.
Let Alice POVM be $\{ V_{0}, V_{1} \}$ (as before)
while by Bob's POVM we denote $\{ W_{0}, W_{1} \}$.
They have pairs of possible local outcomes
$i_{A},i_{B}=0,1$ where $i_{A}$ ($i_{B}$)
corresponds to Alice (Bob) outcome
respectively. Then, performing measurements on
their ancillas they should not only estimate parameters
$p_{0}=Tr(W_{0}^{\dagger}W_{0}\varrho_{A})$,
$q_{0}=Tr(W_{0}^{\dagger}W_{0}\varrho_{B})$
which corresponds to normalised
numbers of outcomes $i_{A}=0$, $i_{B}=0$ respectively.
In addition they should also communicate
classically and count all
cases when they got the results $i_{A}=i_{B}=0$
correlated ie. coming from the copy of the
state $\varrho_{AB}$.
Normalising the resulting
number of the cases ie. dividing it by the number of
all measurements they they got joint correlation probability
\begin{equation}
p_{00}=Tr(V_{0} \otimes W_{0} \varrho_{AB})
\end{equation}
of getting the same outcome $i_{A}=i_{B}=0$ on both sides
from the same copy of the state.

The above process is equivalent to estimation of mean values
$\langle \sigma_{z}^{(A)} \rangle$,
$\langle \sigma_{z}^{(B)} \rangle$, and
$\langle \sigma_{z}^{(A)} \otimes \sigma_{z}^{B} \rangle$
on Alice and Bob local ancillas that were needed
to implement the POVM .
Thus the process is virtually identical to
what happens in usual Bell-type inequality
on two spin-$\frac{1}{2}$ particle where
also marginal and correlation probabilities
are determined.
Summarising Alice and Bob need to determine
probabilities $p_{0}, q_{0}$ and $p_{00}$
of standard Pauli $\sigma_{z}$ measurements
on their ancillas. It is easy to see that
from the probabilities they easily get
the needed mean value as follows
\begin{equation}
\langle A \otimes B \rangle_{\varrho_{AB}}=
a_{+}b_{+}p_{00} + a_{-}b_{-} -[a_{+}b_{-}p_{0}+a_{-}b_{+}q_{0}]
\end{equation}
where $b_{\pm}$ are defined with respect to observable
$B$ in full analogy to $a_{\pm}$.
Thus again we have reduced LOCC measurement of $A \otimes B$ to two
binary POVM-s with more careful data analysis
leading not only to binary marginal distributions
(determined by probabilities $p_{0}$, $q_{0}$) but also to
correlation probability $p_{00}$.
Finally let us note that the above reasoning can be generalised to
multipartite LOCC scheme. Then only the proper hierarchy
of correlation probabilities must be taken into account.

{\it Binary Bell-like inequalities .-}
Note that any Bell inequality involving {\it arbitrary}
bounded observables can be reduced
in this way to the ``binary Bell-like inequality''
involving joint probabilities of binary events
(like $p_{0}$, $q_{0}$, $p_{00}$ and their multipartite
analogs). However validity of such inequality assumes validity
of quantum mechanics (quantum interaction).
Whether and when it is possible to remove the latter assumption
in such modified ''binary Bell inequality'' is an open question.
In any case it seems that there is possibility of 
some universal translation of arbitrary Bell inequalities 
into the legitimate binary ones along the above lines. 
{\it Discussion and conclusions .-}
We have discussed the problem of whether
measurement of single observable with many eigenvalues
can be restricted to estimation of single parameter.
We have shown that it is always possible if
(i)the observable is bounded, ie. has upper and lower bounds
on its spectrum
and (ii) one has well controlled interaction with
single qubit ancilla.

We have constructed the corresponding POVM
and pointed out that it can be achieved with only one
additional quantum bit. Namely the estimated parameter
corresponds to probability of getting one outcome of two possible
in measurement on Pauli operator $\sigma_{z}$ on single qubit
ancilla.

We have also considered the issue of detecting partially
known entanglement with minimal number of estimated
parameters in context of Ref. \cite{Guhne}.
In this case it happens that number of local
observables involved in the measurement is
equal to the corresponding binary POVM-s that can supersede them.
The result of POVM-s, however should be used in more detailed way
to get not only marginal (single parameter) binary
statistics $\{p_{0}, 1-p_{0} \}$,
($\{q_{0}, 1-q_{0} \}$) on Alice (Bob) side but also
join correlation probability $p_{00}$.
It can be generalised to multipartite systems and
leads to compression of usual Bell inequalities
into ``binary Bell-like inequalities'' involving only
joint probabilities of binary events.

Let us observe that in CV case the measurement
of general observable is impossible - due to infinite number
of outcomes one can only measure some approximated
(``digitalised'') observable instead of the original one.
The present binary POVM method seem to
be {\it the  only one} that provides the mean value of
the observable itself rather than its approximation.
The present result has some similarity to recent interferometric
method \cite{Estimator} where final estimation comes form measurement
of Pauli matrix $\sigma_{z}$. However its fundamental difference
can be seen easily when one realises that the
interferometric approach by no means can work
for infinite-dimensional scenario called continuous
variables (CV) case. The observable is there encoded by affine
transformation $A \rightarrow \varrho_{A}=\alpha I + \beta A$
where $I$ stands for identity operator. Hence for  CV
$\varrho_{A}$ is no longer a quantum state
(as the interferometric method of \cite{Estimator} requires)
because has no finite trace.
There is no problems like that for the present method.
Even the presence of square root in formulas defining $V_{0}$, $V_{1}$
does not mean that the discussed difference is equivalent
to that between probabilities and amplitudes in quantum theory.
There is a deeper reason: in the present method the
observable $A$ is encoded directly {\it in global dynamics}
(ancilla-system interaction Hamiltonian that can be inferred
from the POVM) rather than  ``programmed'' into the
``static'' ancilla as it was proposed in Ref. \cite{Estimator}.

There is, however, some important point that
links the present approach with that of Ref. \cite{Estimator}.
Let us recall that some kind of  ``quantum programs''
that implement some physical observable in physical system (ancilla)
has been already presented in previous approach \cite{Estimator}.
Moreover there is  general idea of quantum programming
with quantum data structure \cite{PHAE,Estimator}
and systematic way of quantum gates programming \cite{gates}.

In the above context an intriguing question arises naturally:
is it possible to implement given observable as a kind of
``program'', and if so, what is the most optimal way to do that ?
Form the present analysis we already know that some CV observables
can not be realised as a sort of ``program'' (state of the ancilla).
But one can imagine the scenario: where observable parameters are
``split'' into the programmable part and the one that
is non programmable but can be encoded into dynamics.
In this context one would need measures that would quantify both
parts. It seems that to characterise the second part the
entangling power concept can be important \cite{entanglingpower}
as well as quantum gates programming \cite{gates} and gates cost
\cite{cost}. Also, in case of continuous variables,
the concept of {\it both classical and quantum complexity}
of observable parameters will have to be taken into account.

Finally, it may be interesting to consider application
of the present result in context of Bell inequalities
tests for continuous variables systems \cite{CVBell}.

The author thanks A. Ekert, D. Oi, C. M. Alves, Ch. Fuchs,
M. Horodecki and R. Horodecki for interesting discussions.
The work is partially supported by the project EQUIP,
contract No. IST-1999-11053.

\end{document}